\documentclass[preprint,showpacs,preprintnumbers,amsmath,amssymb]{revtex4}

\usepackage{graphicx}% Include figure files
\usepackage{dcolumn}% Align table columns on decimal point
\usepackage{bm}% bold math

\begin{document}

\preprint{APS/123-QED}

\title{Field-induced Incommensurate Phase in the Strong-Rung Spin Ladder with Ferromagnetic Legs}% Force line breaks with \\

\author{H. Yamaguchi$^{1}$, H. Miyagai$^{1}$, M. Yoshida$^{2}$, M. Takigawa$^{2}$, K. Iwase$^{1}$, T. Ono$^{1}$, N. Kase$^{2,*}$, K. Araki$^{2}$, S. Kittaka$^{2}$, T. Sakakibara$^{2}$, T. Shimokawa$^{3}$, T. Okubo$^{2}$, K. Okunishi$^{4}$, A. Matsuo$^{2}$, and Y. Hosokoshi$^{1}$}
%\altaffiliation[Present address]{Physics Department, XYZ University.}%Lines break automatically or can be forced with \\

%\author{Second Author}
\email {yamaguchi@p.s.osakafu-u.ac.jp}

\affiliation{
$^1$Department of Physical Science, Osaka Prefecture University, Osaka 599-8531, Japan \\ 
$^2$Institute for Solid State Physics, The University of Tokyo, Chiba 277-8581, Japan\\
$^3$Department of Earth and Space Science, Faculty of Science, Osaka University, Osaka 560-0043, Japan\\
$^4$Department of Physics, Niigata University, Niigata 950-2181, Japan\\}

\date{\today}% It is always \today, today,

\begin{abstract}
We report magnetization, specific heat, and NMR measurements of 3-Br-4-F-V [= 3-(3-bromo-4-fluorophenyl)-1,5-diphenylverdazyl], strong-rung $S$=1/2 Heisenberg spin ladder with ferromagnetic leg interactions. 
We explain the magnetic and thermodynamic properties based on the strong-rung regime.
Furthermore, we find a field-induced successive phase transition in the specific heat and the nuclear spin-lattice relaxation rate 1/$T_{1}$.
$^{19}$F-NMR spectra for higher and lower temperature phases indicate partial magnetic order and incommensurate long-range order, respectively, evidencing the presence of frustration due to weak interladder couplings.
\end{abstract}

                 
\maketitle
The spin ladder is prototypical quantum spin system.
The $S$ = 1/2 antiferromagnetic (AFM) spin ladder, which exhibits AFM rung and leg interactions, has been extensively studied in relation to field-induced quantum phase transitions and high-$T_c$ superconductors~\cite{SL1,SLSC}. 
Recent investigations on new copper complexes, (C$_5$H$_{12}$N)$_2$CuBr$_4$ (BPCB) and (C$_7$H$_{10}$N)$_2$CuBr$_4$ (DIMPY), have extended our understanding of $S$ = 1/2 AFM spin ladders and their field-induced quantum liquid phases~\cite{BP_CM,DY_CM,DY_M,BP_NMR,DY_NMR}.
BPCB and DIMPY exhibit strong-rung and strong-leg coupling regimes, respectively, and indicate several different magnetic and thermodynamic properties.
The interesting contrast in the associated Luttinger liquids (LLs) for strong-rung and strong-leg coupling regimes has received particular attention~\cite{BP_NMR,DY_NMR} because these two regimes represent model systems of spinless fermions with repulsive and attractive interactions, respectively.

In contrast, ferromagnetic (FM) chain-based spin ladders, where FM chains are coupled by AFM rung interactions, behave very differently than AFM spin ladders.
FM chains tend to suppress quantum fluctuations, making the stabilization of quantum states uncertain~\cite{1DFMth_1,1DFMth_2}.
Since FM chain-based systems are expected to have relatively larger nonlinear terms in their low-energy dispersion relations from 1D FM interactions, LL formation allows one to examine the limitations of conventional linear LL theory~\cite{nonLL1, nonLL2}.
The phase diagram of FM chain-based spin ladder is parameterized by the ratio between the rung and the leg interactions, $\gamma =|J_{\rm{rung}}/J_{\rm{leg}}|$, and XXZ-type exchange anisotropy~\cite{FL1,FL3,FL4,FL5}.
For the isotropic case, a fully gapped rung-singlet (RS) state is stabilized. 
Vekua \textit{et al.} argued that the LL regime can appear when a magnetic field is applied~\cite{FL4}.

We previously reported the first experimental realization of an $S$ = 1/2 FM chain-based spin ladder in 3-Cl-4-F-V  [= 3-(3-chloro-4-fluorophenyl)-1,5-diphenylverdazyl]~\cite{3Cl4FV}, where $\gamma$ was evaluated as 0.55. 
However, in zero magnetic field the energy gap disappears and phase transitions to an ordered state are observed.
Thus, the AFM rung coupling with $\gamma$ = 0.55 is considered too weak to prevent magnetic order and stabilize quantum phases.
Because 3-Cl-4-F-V has strong-leg coupling, a model substance that has a strong-rung coupling with $\gamma \textgreater 0.55$ is required to examine quantum phases and the differences between coupling regimes.
More recently, we have reported a new model material for an $S$ = 1/2 spin ladder with FM leg interactions, 3-Br-4-F-V [= 3-(3-bromo-4-fluorophenyl)-1,5-diphenylverdazyl], where the Cl in 3-Cl-4-F-V is replaced by Br~\cite{3ladders}.
The low-temperature magnetization data are well described within the strong-rung regime with $\gamma$ = 1.5~\cite{3ladders}.

%要約
In this Letter, we report magnetization, specific heat, and NMR measurements of 3-Br-4-F-V. 
We successfully explain the magnetic and thermodynamic properties within the expected spin-ladder model with $\gamma$ = 1.5. Furthermore, we observe a field-induced successive phase transition, which occurs just below the LL regime.
The NMR spectra in the field-induced phases indicate an incommensurate spin structure, demonstrating the presence of frustration in the weak interladder couplings.

%%(実験方法) 
Single crystals of 3-Br-4-F-V were synthesized as described in Ref.~\cite{3ladders}.
The magnetic susceptibility and magnetization curves were measured using a capacitive Faraday magnetometer in a dilution refrigerator.
The specific heat was measured using a hand-made apparatus by a standard adiabatic heat-pulse method for 0.35 K $<T<$ 2.0 K. 
Using specific heat measurements, we confirmed that no nontrivial magnetic anisotropy exists for $H\parallel$ $a$ and $H\perp$ $a$ and so all experiments were performed for $H\perp$ $a$.
This isotropy is due to the nature of organic radical systems.
NMR measurements were performed using a single crystal with approximate dimensions of 3.5$\times$1.0$\times$0.5 mm$^3$.
An external magnetic field was applied along the $ac$ direction in order to make all F sites physically equivalent.
The $^{19}$F-NMR spectra were obtained with a fixed magnetic field by summing the Fourier transform of the spin-echo signal at equally spaced rf-frequencies or by recording the integrated intensity of the spin-echo signal at discrete frequencies .
We determined 1/$T_{1}$ by fitting the spin-echo intensity $M$($t$) as a function of the time $t$ after the inversion pulse to the stretched exponential recovery function $M$($t$) = $M_{\rm{eq}}$ $-$ $M_{0}$exp${\{}$$-$($t$/$T_1$)$^{\beta}{\}}$,
where $M_{\rm{eq}}$ is the intensity at thermal equilibrium and ${\beta}$ is the stretch exponent that provides a measure of inhomogeneous distribution of 1/$T_{1}$. When 1/$T_{1}$ is homogeneous, the value of ${\beta}$ is close to one.
When the spectra became broad in the AFM ordered state, 1/$T_{1}$ was measured at the spectral center.

%%既知物性
Figures 1(a) and 1(b) show the molecular structure of 3-Br-4-F-V and its spin-density distribution, respectively.
$Ab$ $initio$ molecular orbital (MO) calculations revealed that about 63 ${\%}$ of the total spin-density is present on the verdazyl ring including four nitrogen atoms and each of the upper two phenyl rings also have a relatively large spin-density of about 15 ${\%}$~\cite{MO}.
Considering such a spin-density distribution, the column structure along the $a$-axis should be regarded to form uniform spin chains.
A short contact connecting two neighboring chains makes a two-leg ladder structure, as shown in Fig. 1(c). 
From our previous analysis of the magnetization, exchange interactions are estimated as $J_{\rm{rung}}/k_{\rm{B}}$ = 12.5 K and $J_{\rm{leg}}/k_{\rm{B}}$ = $-$8.3 K ($\gamma$ = 1.5)~\cite{3ladders}, defined within the Heisenberg spin Hamiltonian:
\begin{equation}
\mathcal {H} = J_{\rm{leg}}{\sum^{}_{ij}}\textbf{{\textit S}}_{i,j}{\cdot}\textbf{{\textit S}}_{i+1,j}+J_{\rm{rung}}{\sum^{}_{i}}\textbf{{\textit S}}_{i,1}{\cdot}\textbf{{\textit S}}_{i,2}-g{\mu _B}H{\sum^{}_{ij}}\textbf{{\textit S}}_{i,j},
\end{equation}
where $\textbf{{\textit S}}_{i,j}$ are $S$=1/2 spin operators acting on site $i$ of leg $j$ = 1,2 of the ladder, the $g$-factor is 2.00, ${\mu}$$_B$ the Bohr magneton, and $H$ the external magnetic field. 
The ground state for this model is the RS with an excitation gap $\Delta \approx  6.9$\,K to the lowest triplet state.
 
%We calculated the magnetic susceptibility as a function of $\gamma$ assuming an $S$ = 1/2 two-leg spin-ladder with FM leg interactions by using the quantum Monte Carlo (QMC) method~\cite{QMC}. 
%Although there is no distinct $\gamma$ dependence of $\chi$, ${\chi}T$ is very sensitive to changes in $\gamma$. 
%There are also weak inter-ladder couplings with a magnitude something below 1 K which cannot be reliably determined by the present methods~\cite{MOseido}.

%%(相図に関して、低温磁化率、磁化曲線、比熱)
Figures 2(a) and 2(b) show the low-temperature behavior of $\chi$ ($\chi$ = $M/H$) and the temperature derivative d$\chi$/d$T$, respectively, under various magnetic fields. 
We observe an extremum in $\chi$, which changes from a minimum at low fields to a maximum at high fields around 7.0 T, as shown in Fig. 2(a). 
Many theoretical studies of 1D gapped spin systems predict extremum in $\chi$ at the crossover temperature to the LL regime~\cite{the1,the2,the3} and experimentally observed in BPCB and DIMPY~\cite{BP_CM, DY_M}.
Our experimental results clearly demonstrate the predicted crossover behavior.
We plot these specific temperatures in the field-temperature phase diagram, as shown in Fig. 2(f).
The obtained crossover line well reproduces predicted dome-like behavior with a deviation in higher-field region~\cite{the2,divTLL}.

Furthermore, d$\chi$/d$T$ exhibits a distinct sharp peak at a temperature below the extremum of $\chi$, as shown by the arrows in Figs. 2(b).
These peaks must indicate phase transition to the 3D ordered phase, which is also predicted on 1D gapped spin systems with weak three-dimensional (3D) couplings~\cite{the2}.
We calculated magnetic susceptibilities using the quantum Monte Carlo (QMC) method~\cite{QMC} with the same parameters as in Ref.~\cite{3ladders} and reproduced the broad extrema, as shown in Fig. 2(a).
%磁化曲線に関して
Figures 2(c) and 2(d) show magnetization curves and their field derivatives (d$M$/d$B$) at various temperatures, respectively. 
The magnetization curves exhibit an excitation gap of about 5.1 T.
We have calculated the magnetization curves using the finite-temperature density matrix renormalization group (DMRG) method~\cite{DMRG} and obtained good agreement with experimental results, as shown in Fig. 2(c).
The d$M$/d$B$ values indicate asymmetric double peak structures, as shown in Fig. 2(d). 
We plot these peak magnetic fields in the field-temperature phase diagram, as shown in  Fig. 2(f).
The observed peak fields are consistent with the phase boundaries determined from other experimental results.

%比熱に関して%
Figure 2(e) shows the temperature dependences of the magnetic specific heat $C_{\rm{p}}$.
The lattice contribution is subtracted from the total specific heat assuming the Debye's $T^3$-law as $0.015T^3$ (J/mol K).
We find distinct double peak structures in the field-induced phase, which resembles the phase transition accompanied by a double peak in the specific heat for 3-Cl-4-F-V~\cite{3Cl4FV}.
The high temperature peaks are located slightly lower than the broad extrema in $\chi$, while the low temperature peaks agree well with the phase boundary evaluated from the peaks in $d\chi/dT$, as shown in the phase diagram in Fig. 2(f).
We calculated the magnetic specific heat using the QMC method and reproduced the broad peak near 3.5 K associated with an excitation gap of the RS in the gapped phase and the high-temperature regions in the gapless phase, as shown in Fig. 2(e). 
Calculations of $C_{\rm{m}}$/$T$ at the temperatures corresponding to the field-induced gapless phase tend toward a constant value at low temperatures, corresponding to the $T$-linear behavior of $C_{\rm{m}}$ expected in LL regime.
Conversely, the experimental results indicate a rapid increase with decreasing temperatures in corresponding temperature regions owing to the peaks associated with the phase transition induced by finite interladder couplings. 
This phase transition occurs immediately below the crossover to the LL regime, which should be attributed to the development of correlations length in the LL regime. 
The observed double peak structure indicates a successive phase transition.
The possible origin is partial order of magnetic moment, in which the specific components of the magnetic moment order in a stepwise fashion, as expected in 3-Cl-4-F-V~\cite{3Cl4FV}.
Such behavior is often induced by large magnetic anisotropy and/or a noncollinear magnetic structure.

%NMR T1
In order to further examine the field-induced phases, we carried out $^{19}$F-NMR measurements down to about 0.1 K.
At low temperatures, the nuclear spin-lattice relaxation rate $T_1^{-1}$ exclusively probes the spin dynamics on the radical site, corresponding to the long time behavior of local spin-spin correlation functions.
Figure 3(a) shows the temperature dependence of $T_1^{-1}$ at various magnetic fields.
In the RS regime at 1.5 and 4.0 T, $T_1^{-1}$ rapidly decreases with decreasing temperature below 8 K because of the energy gap. 
In the gapless phase above 6.0 T, $T_1^{-1}$ rapidly increases with decreasing temperature below 1 K and shows a distinct sharp peak for each field, indicating a magnetic phase transition with critical slowing down.  
Here, the recovery curves fit well to a single exponential function ($\beta$ = 1) above about 0.8 K, indicating homogeneous nuclear relaxation.
Below 0.8 K, the stretch exponent $\beta$ decreased monotonically with decreasing temperature and reached 0.6 at 0.2 K probably due to the distribution of the AFM internal field within the frequency window of $\pm$ 0.2 MHz covered by the exciting rf pulse.
As shown in the inset of Fig. 3(a), $T_1^{-1}$ also shows a small shoulder at a higher temperature than the peak for each field.
Those peak and shoulder positions are consistent with the phase boundaries evaluated from other experimental results, as shown in Fig. 2(f). 
Thus, we conclude that the temperature dependence of $T_1^{-1}$ supports the successive phase transitions expected from the magnetic specific heat.
At 5.0 T near the critical field, $T_1^{-1}$ shows a minimum around 1 K and then slightly increases with decreasing temperature.
This increase is consistent with the power law behavior $T_1^{-1}\propto T^{-0.5}$ as one might expect in the quantum critical regime~\cite{T105}.
However, $T_1^{-1}$ suddenly decreases below about 0.17 K because of the phase transition to a magnetically ordered state.
Further study is needed to explore the quantum critical point associated with spinless fermions.

%NMR spectl
Here, we discuss the spin structures in the field-induced phases from the $^{19}$F-NMR spectra.
Figure 3(b) shows the temperature dependence of the $^{19}$F-NMR spectra at 7 T. 
The spectra above 1.2 K show paramagnetic single lines. 
In the lower temperature phase (0.24 K and 0.62 K), the paramagnetic component completely disappears, and a double-horn type line shape with a continuum of finite intensity between the two peaks appears, indicating a 3D ordered state.
Such a line shape provides definitive evidence for the formation of an incommensurate spin structure.      
The spectra in the higher temperature phase (0.70 K and 0.77 K) seem to consist of two components; sharp line corresponding to a paramagnetic phase and broad line associated with the incommensurate spin structure.
We can consider two cases, a partial order or macroscopic phase separation.
Our results support the former case.
The two distinct phase transitions observed in $T_1^{-1}$ cannot be explained by simple coexistence of paramagnetic and ordered phases.
Because $T_1^{-1}$ was measured at the peak position of the spectra above 0.7 K, it should be dominated by the paramagnetic-like component in case of macroscopic phase separation. 
Nevertheless $T_1^{-1}$ at 7.0 T shows the clear shoulder at about 0.8 K, indicating that the spin dynamics of the paramagnetic-like component are definitely affected by the AFM ordered moment through the short-range hyperfine interaction.
Such behavior can be understood only by partial but homogeneous magnetic order.
Furthermore, if the macroscopic phase separation is realized, one can expect that the recovery curve for $T_1^{-1}$ consists of two components; one is a fast recovery from the paramagnetic phase and the other is a slow one from the ordered phase.
However, the recovery curve in the higher temperature phase can be fitted by a stretched exponential function with a relatively large $\beta$$\sim$0.9 as shown in the left inset of Fig. 3(a), which indicates that the difference between $T_1^{-1}$ of the paramagnetic- and ordered-like components is small. 
This result can also be explained by the partial order scenario, in which the $^{19}$F-NMR nuclear is affected by both the paramagnetic and the ordered moments.

%梯子間の相互作用
Finally, we take into account of the interladder couplings in order to understand the origin of the incommensurate spin structure.
Because the field and temperature dependences of magnetization are well reproduced within a spin ladder model even in the ordered phase, the values of the interladder interactions are expected to be quite small, resulting in the formation of very longwave incommensurate structure.
In fact, the MO calculations indicate that there are three kinds of possible small interladder interactions with absolute values less than about 0.5 K~\cite{3ladders}.
These interactions can result in two types of triangular unit on the lattice, inducing frustration through the combination of their signs.
One consists of only interladder interactions, and the other consists of leg and interladder interactions.
However, since these conclusions strongly depend on the calculation method~\cite{MOseido}, the small values of the interladder interactions are not reliable.    
Further investigation by neutron scattering will yield quantitative information about the incommensurate propagation vector and 
the dispersion relations and so define the dominant interladder interactions causing the frustration.

In summary, we measured the magnetization, specific heat, and NMR of 3-Br-4-F-V, a strong-rung $S$=1/2 Heisenberg spin ladder with FM leg interactions.
We have explained the magnetic and thermodynamic properties within the strong-rung regime with $\gamma =|J_{\rm{rung}}/J_{\rm{leg}}|$ = 1.5 using QMC and DMRG calculations.
The magnetic susceptibility demonstrates the predicted behavior at the crossover to the LL regime. 
Furthermore, we observe two field-induced successive phase transitions immediately below the crossover temperature. 
The NMR spectra indicate two distinct field-induced phases.
A partial order of the magnetic moments occurs in the high temperature phase, while in the low temperature phase all spins participate in an incommensurate ordered state.
This provides evidence for frustration caused by the weak interladder couplings.
Hence, this material will stimulate investigation into the intrinsic physics of FM chain-based quantum spin systems and low-dimensional systems weakly coupled by frustrated interactions.

We thank T. Tonegawa for the discussions. This research was partly supported by KAKENHI (Nos. 24740241, 24540347, and 24340075) and the CASIO Science Promotion Foundation．A part of this work was performed under the interuniversity cooperative research program of the joint-research program of ISSP, the University of Tokyo. Some computations were performed using the facilities of the Supercomputer Center, the ISSP, The University of Tokyo.

\begin{figure}[t]
\begin{center}
\includegraphics[width=17pc]{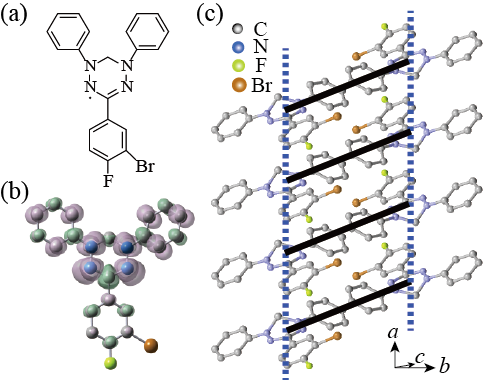}
\caption{(color online) (a) The moleculer structure of 3-Br-4-F-V and (b) its spin density distribution. Purple and green shapes correspond to positive and negative spin densities, respectively. The isodensity surface corresponds to a cutoff value of 0.001 e bohr$^{-3}$. (c) Crystal structures of 3-Br-4-F-V forming the spin ladder. Solid and dashed lines stand for $J_{\rm{rung}}$ and $J_{\rm{leg}}$, respectively.}\label{f1}
\end{center}
\end{figure}

\begin{figure}[t]
\begin{center}
\includegraphics[width=38pc]{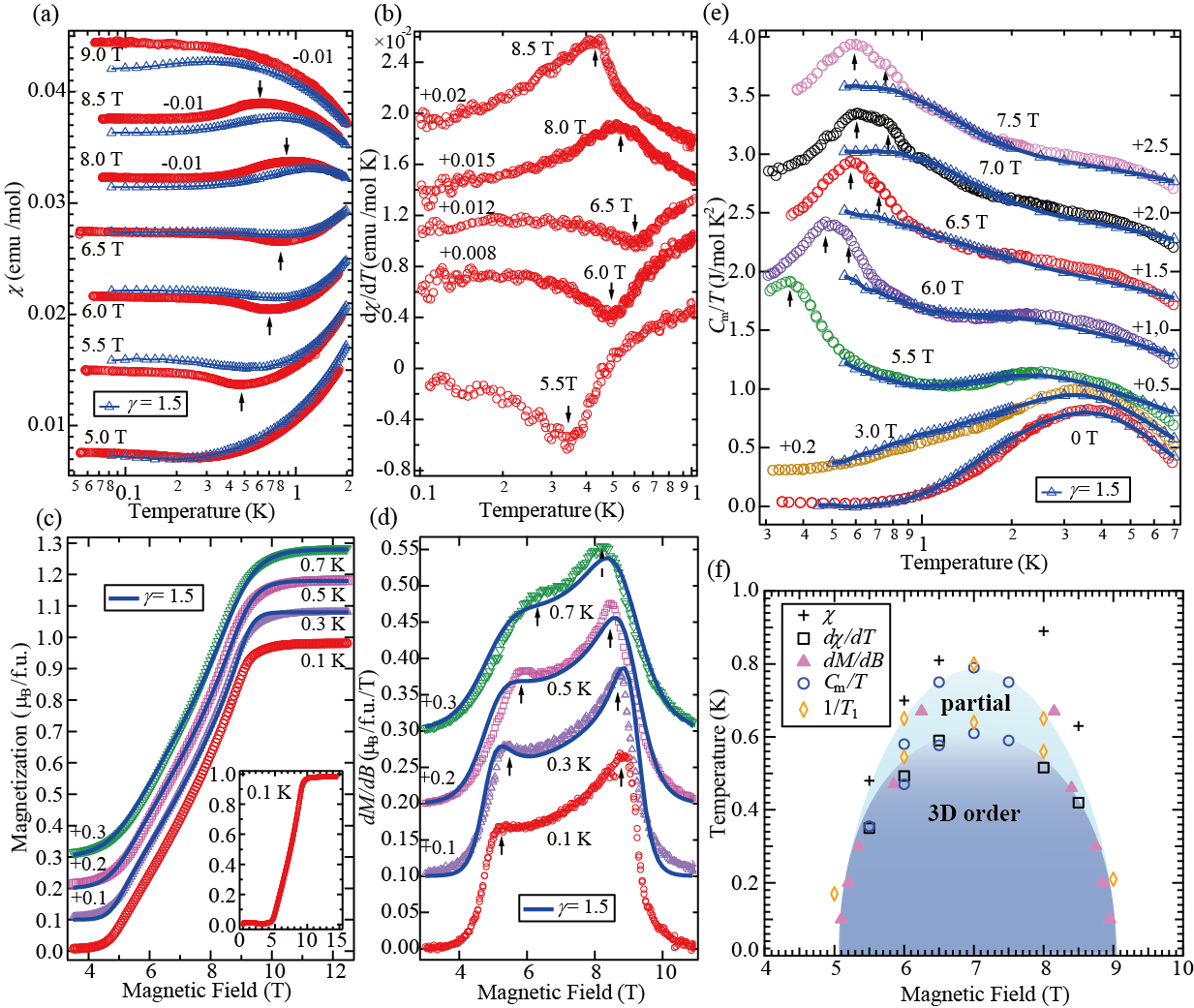}
\caption{(color online) Temperature dependence of magnetic susceptibility $\chi$ (a) and its temperature derivative d$\chi$/d$T$ (b). Magnetization curve (c) and its field derivative d$M$/d$B$ (d).
The inset shows the magnetization curve at 0.1 K for the full field range.
(e) Temperature dependence of $C_{\rm{m}}/T$.
(f) Magnetic field versus temperature phase diagram, showing partially ordered and incommensurate 3D ordered phases. 
For clarity, the values of the vertical axes have been shifted arbitrarily in all experimental results, and the shifted values are shown in each figure. 
All calculations are performed for $\gamma$ = $|J_{\rm{rung}}/J_{\rm{leg}}|$ = 1.5. and indicated by the blue lines and open triangles.}\label{f2}
\end{center}
\end{figure}

 \begin{figure}[t]
\begin{center}
\includegraphics[width=18pc]{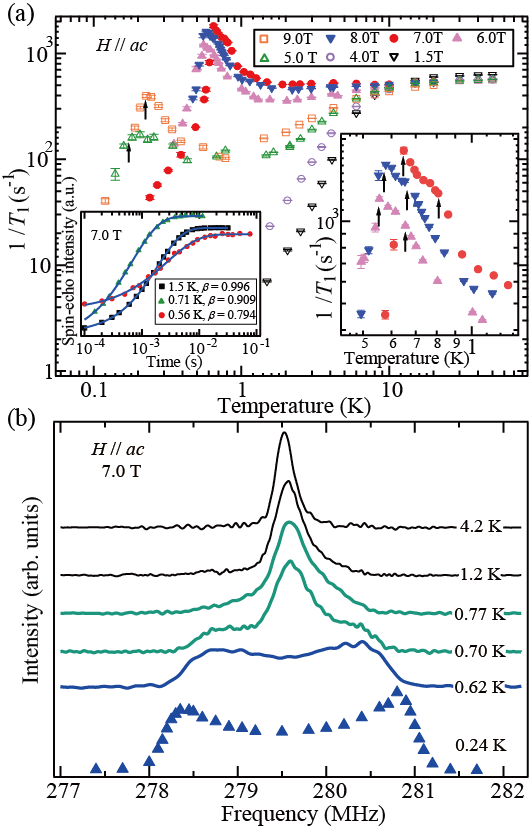}
\caption{(color online) (a) Temperature dependence of $T_1^{-1}$ for $^{19}$F nuclei at various magnetic fields. The error bar is the s.d. in fitting the recovery curve. The right inset shows the extended temperature region associated with the phase transitions. The left inset shows the recovery curves for $T_1^{-1}$ at 7.0 T fitted with the stretched exponential functions indicated by the solid curves.
(b) Temperature dependence of $^{19}$F-NMR spectra at 7.0 T for various temperatures.}\label{f1}
\end{center}
\end{figure}


\begin{thebibliography}{99}
\bibitem{MO}
The $ab$ $initio$ MO calculations were performed using the UB3LYP method as broken-symmetry hybrid density functional theory calculations.
All the calculations were performed using the Gaussian 09 program package and the 6-31G basis sets.
\bibitem{QMC} 
Our QMC code is based on the continuous-imaginary-time-loop algorithm~\cite{QMC1} and the directed loop algorithm in the stochastic series expansion representation~\cite{QMC2}. 
The calculation was performed for a system size of 96 under a periodic boundary condition using the ALPS application~\cite{ALPS}.
\bibitem{DMRG}
Finite-temperature DMRG calculations~\cite{DMRG1,DMRG2,DMRG3} were performed for the quantum transfer matrix under the periodic boundary condition in the imaginary time direction. 
The maximum number of the basis kept is 200, within which the computational results were well converged.
\end{thebibliography}
\end{document}